\documentclass[prl,twocolumn,preprintnumbers,amsmath,amssymb,letterpaper]{revtex4}

\usepackage{graphicx}
\DeclareGraphicsExtensions{.eps}

\begin{document}

\preprint{cond-mat/0411724}
 
\title{How the Liquid-Liquid Transition Affects Hydrophobic Hydration in
Deeply Supercooled Water}

\author{Dietmar Paschek} 
\email{dietmar.paschek@udo.edu}
\affiliation{Physikalische Chemie, Universit\"at Dortmund, 
  Otto-Hahn-Str. 6, D-44221 Dortmund, Germany}


\date{\today}
\begin{abstract}
We determine the phase diagram of liquid supercooled water by extensive computer
simulations using the TIP5P-E model [J. Chem. Phys. {\bf 120}, 6085 (2004)].
We find that the transformation of water into a low density liquid 
in the supercooled range strongly enhances the solubility of
hydrophobic particles. The transformation of water into 
a tetrahedrally structured liquid is accompanied
by a minimum in the hydration entropy and enthalpy. The corresponding change
in sign of the solvation heat capacity indicates a loss of one characteristic
signature of hydrophobic hydration. The observed behavior 
is found to be qualitatively in accordance 
with the predictions of the information theory model of Garde et al.
[Phys. Rev. Lett. {\bf 77}, 4966 (1996)].
\end{abstract}

\maketitle

The thermodynamical anomalies of liquid water are considered to be
caused by a transformation
between two different liquid forms of water buried in the 
deeply supercooled region \cite{Debenedetti:2003}.
The two differently dense liquids have well characterized
counterparts in the glassy state: The 
(very) high density and low density amorphous ice forms \cite{Mishima:84,Loerting:2001}.
Computer simulation studies have furnished a picture of
a first order liquid-liquid phase transition between two
liquids ending up in a metastable critical point
\cite{Poole:92,Tanaka:96:1}. Although singularity free scenarios 
might as well explain the properties of supercooled water \cite{Debenedetti:2003}, 
there is experimental support for the liquid-liquid critical
point hypothesis from the changing slope of
the metastable melting curves 
observed for different ice polymorphs \cite{Mishima:1998:2,Mishima:2000}.
To make the situation even more puzzling, recent computer simulations
provide evidence that there might be even more than 
one liquid-liquid transition \cite{Brovchenko:2003}.

One prominent anomaly of liquid water is the increasing
solubility of hydrophobic gases with decreasing temperature
\cite{Wilhelm:77}. This behavior is the consequence
of a negative solvation entropy of small hydrophobic particles
\cite{PrattRev:2002}. 
Since the corresponding solvation enthalpy is also
negative, the observed low solubility
(the large positive solvation free energy) of small hydrophobic
particles is due to a dominating entropy effect. 
The origin of the negative hydration entropy 
is widely regarded as being due to the bias in the hydration waters
orientational space, as the water molecules are trying to preserve
their hydrogen bond network \cite{Southall:2002}.
Entropy and enthalpy effects have been shown to be
determined by the water in the first hydration shell \cite{Paschek:2004:2}. 
In addition, the broken hydrogen bond states, with hydrogen bond donors/acceptors pointing towards
the hydrophobic particles, are increasingly populated with increasing
temperature \cite{Southall:2002}. Hence,
the {\em positive} solvation heat capacity of
hydrophobic particles is considered as one of the key signatures for
hydrophobic hydration \cite{Silverstein:2000}.
A recent experimental study of Souda \cite{Souda:2004}, investigating 
alkane layers adsorbed on an amorphous
solid water substrate, shows that the alkane phase gets soaked into
the water phase in the temperature region close to the suspected
glass transition temperature \cite{Mishima:2004}. One possible explanation
(among others) might be that the highly viscous low density liquid form of
water provides a significantly increased solubility for hydrophobic 
molecules. The scope of the present contribution is to monitor hydration of
a small hydrophobic particle as we penetrate the deeply 
supercooled region. Our particular interest is to elucidate
how the transformation into a 
highly tetrahedrally ordered
low density liquid (LDL) phase \cite{Paschek:99} affects 
hydrophobic hydration.

An important problem regarding simulations of highly viscous liquids
close to the glass transition is to provide proper sampling. 
To overcome this problem we perform parallel tempering simulations 
of an extended ensemble of states \cite{FrenkelSmit} using 
the technique of volume-temperature
replica exchange molecular dynamics simulation (VTREMD) \cite{Paschek:2004:3}.
To represent liquid supercooled water, we employ the
 recently proposed TIP5P-E model for water, which was fitted to reproduce
 waters  density maximum and related
thermodynamical anomalies while
treating the Coulomb interactions with the Ewald sum \cite{Rick:2004}.
For the VTREMD simulation \cite{VTREMD} we consider a 
grid of 440 ($V$,$T$)-states \cite{STATES}.
\begin{figure}[!t]
  \centering
  \includegraphics[angle=0,width=7cm]{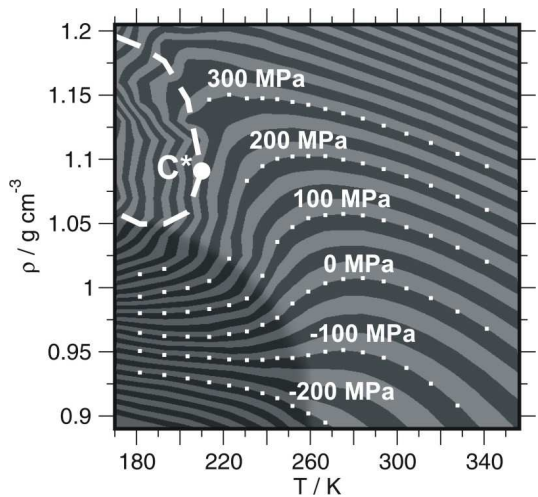}

  \caption{\small Phase diagram for stable and supercooled liquid
    TIP5P-E water.
    The spacing between
    different contour colours corresponds to a pressure drop of 25 MPa. 
    Selected isobars are indicated. 
    The heavy dashed line denotes
    the
    high density / low density liquid
    (HDL/LDL) coexistence line.
    The HDL/LDL critical point is located
    at $C^*$ with $T^*\!=\!210\,\mbox{K}$, $P^*\!=\!310\,\mbox{MPa}$ and
    $\rho^*\!=\!1.09\,\mbox{g}\,\mbox{cm}^{-3}$. Darker shading indicates the
    LDL-basin.
  }
  \label{fig:01}
\end{figure}
Starting from a set of equilibrated initial configurations obtained at
ambient conditions, 
the VTREMD simulation \cite{techdetails} was conducted for $20\,\mbox{ns}$,
providing a total $8.8\,\mu\mbox{s}$ worth of trajectory data.
The average time interval between two successful 
state-exchanges was obtained to be about $3\,\mbox{ps}$. During the entire
course of the simulation each replica
has crossed the whole temperature and density interval several times.
After an initial equilibration period of about $4\,\mbox{ns}$, 
the average pressure
and potential energies show convergence even for the lowest temperatures
and the remaining  $16\,\mbox{ns}$ are used for analysis.

\begin{figure}[!t]
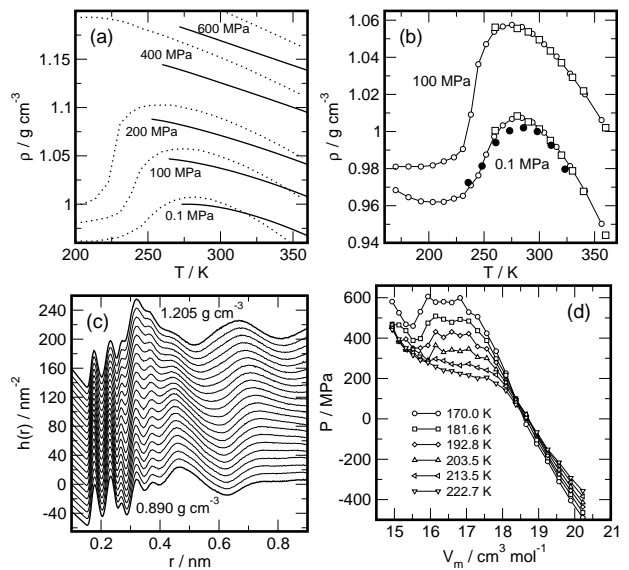

  \centering
  \includegraphics[angle=0,width=4cm]{fig2a}
  \includegraphics[angle=0,width=4cm]{fig2b}

  \includegraphics[angle=0,width=4cm]{fig2c}
  \includegraphics[angle=0,width=4cm]{fig2d}
  \caption{\small (a): Selected isobars for TIP5P-E water (dotted lines) 
    compared with
    experimental data \cite{Wagner:2002}. (b) Two isobars for
    TIP5P-E. Open circles: Interpolated data
    from the VTREMD $P(\rho,T)$
    data-set. Open squares: Constant
    temperature/pressure (NPT) MD simulations.
    Filled circles: Ref. \cite{Rick:2004}. 
    (c) 
    $h(r)\!=\!4\pi r (N/V)
    \left[ 
      0.092\,g_{\mbox{OO}}(r)
      +0.486\,g_{\mbox{HH}}(r)
      +0.422\,g_{\mbox{OH}}(r)-1
    \right]
      $-functions for the $222.7\,\mbox{K}$ isotherm.
    The functions are shifted by an increment of $10\,\mbox{nm}^{-2}$.
    (d) Near- and sub-critical isotherms.  
  }
  \label{fig:02}
\end{figure}

Figure 1 shows the phase diagram of liquid supercooled water 
in terms of a contour plot of the $P(\rho,T)$-data 
as obtained by the VTREMD simulations. 
The TIP5P-E phase diagram apparently exhibits a first
order phase transition between two metastable liquid phases, ending
in a second critical point $C^*$. The location of the coexistence line,
estimated from a Maxwell construction using the subcritical isotherms
shown in Figure \ref{fig:02}d, 
has to be seen as a rough guess only, since the exact form of the
van der Waals loops in the two phase region might depend on the 
system size. The location of $C^*$ 
is close the values reported by Yamada et al. for the original 
TIP5P model 
with $T^*\!=\!217\,\mbox{K}$, $P^*\!=\!340\,\mbox{MPa}$ and
    $\rho^*\!=\!1.13\,\mbox{g}\,\mbox{cm}^{-3}$ \cite{Yamada:2002}.
Figure 2a compares 
experimental data \cite{Wagner:2002} with
several selected isobars according to a linear
interpolation from the TIP5P-E $P(\rho,T)$ data-set.
Although the location of the density maximum at normal pressure is
close to the experimental values, 
the TIP5P-E model exhibits a considerably larger thermal expansivity
as real water and is significantly more compressible.
Figure 2b shows that the isobars derived from VTREMD data-set
match exactly the data obtained from conventional $NPT$-simulations. 
For comparison the isobar according 
to Rick's simulation is shown \cite{Rick:2004}.
We denote small, but significant differences. Since both 
simulations were using Ewald summation, 
their origin is not quite clear at present. 
We would like to emphasize one particular detail of the low pressure isobars,
indicated in Figures 1 and 2b: The density goes through 
a minimum after the transformation into the low density liquid has taken
place. Apparently at low temperatures the expansivity seems to behave as 
it would be expected for a conventional liquid. 
I. Brovchenko et al.\ \cite{Brovchenko:2003} and recently P.\ H.\ Poole 
\cite{Poole:private} have made similar observations 
for the ST2-model. Figure 2c shows the evolution of
the composite radial distribution function $h(r)$
along the $222.7\,\mbox{K}$-isotherm. The highlighted patterns obtained for the
lowest and highest densities show strong similarity with the functions
determined experimentally by Bellissent-Funel et al.
for high and low density amorphous ice \cite{Bellissent-Funel:87}.
The observed low density pattern has been shown to be related to a high
tetrahedral order of water's first hydration shell \cite{Paschek:99}.
Similarities between the local structure of LDA and ice $I\mbox{h}$ have 
also been demonstrated recently by Finney et al. from the analysis of 
neutron scattering experiments
\cite{Finney:2002}. 
\begin{figure}[!t]
  \centering
  \includegraphics[angle=0,width=6.0cm]{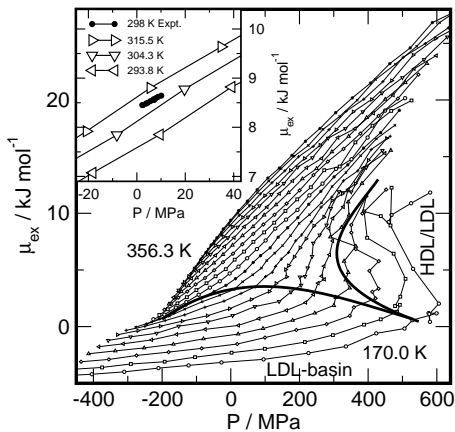}
  \caption{\small 
    Excess chemical potential of Ar dissolved in TIP5P-E water as a
    function of $P$ and $T$. Thin lines indicate all 
    calculated isotherms ranging from $170\,\mbox{K}$
    to $356.3\,\mbox{K}$. The insert shows experimental data for 
    $298\,\mbox{K}$ \cite{Kennan:90} and simulated
    isotherms close-by. Thick lines indicate the LDL-basin
    and the HDL/LDL-coexistence region (see Figure \ref{fig:01}).
  }
  \label{fig:03}
\end{figure}
\begin{figure}[!t]
  \centering
  \includegraphics[angle=0,width=6.0cm]{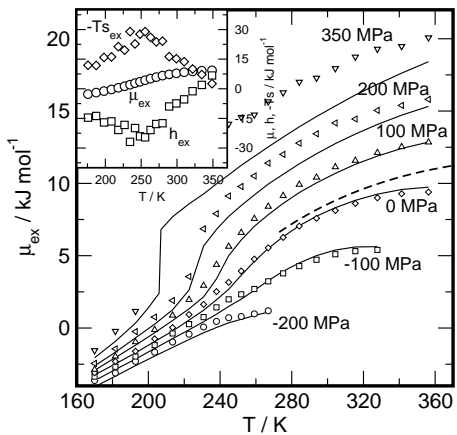}
  \caption{\small 
    The symbols indicate six isobars, interpolated
    from the $\mu_{ex}(\rho,T)$
    TIP5P-E data-set. The dashed line indicates experimental data
    for $0.1\,\mbox{MPa}$ \cite{Prini:89}. Full lines represent
    predictions for
     $\mu_{ex}\!=\!a \rho^2T +b$ 
    with 
    $a\!=\!8\times 10^{-2}\,\mbox{kJ}\,\mbox{mol}^{-1}\,
    \mbox{K}^{-1}\,\mbox{cm}^6\mbox{g}^{-2}$
    and
    $b\!=\!-16\,\mbox{kJ}\,\mbox{mol}^{-1}$.
    The insert shows enthalpy ($h_{ex}$) and entropy ($-T\,s_{ex}$)
    contributions to $\mu_{ex}$ for the 
    $0\,\mbox{MPa}$ isobar.
  }
  \label{fig:04}
\end{figure}
We do not observe indications for the
presence of further liquid-liquid transitions for TIP5P-E, as suggested by Brovchenko et
al. for the ST2 model \cite{Brovchenko:2003}.  However, further phase transitions 
might be present at even lower temperatures. Finally, no signs for
crystallization are observed, as denoted by Yamada et
al. \cite{Yamada:2002}. Probably our smaller system size and the random walk
of the replicas through state-space, regularly reaching stable
regions, prevents the systems from crystallizing.

The hydrophobic hydration of a Lennard-Jones Argon particle
($\sigma_{\mbox{Ar-O}}\!=\!0.3290\,\mbox{nm}$ and
$\epsilon_{\mbox{Ar-O}}/k\!=\!98.9\,\mbox{K}$) \cite{Paschek:2004:1}
 is given as excess chemical potential $\mu_{ex}= -\beta^{-1} \ln \gamma$
for infinite dilution, where $\gamma$ is the solubility and $\beta\!=\!1/kT$. 
We employ the Widom particle insertion method \cite{FrenkelSmit} 
with $\mu_{ex}\;= \; - \beta^{-1}\left< \exp(-\beta\,\Phi) \right>$,
 where $\Phi$ is the energy of an inserted test-particle. The brackets
$\left< \ldots\right>$ indicate canonical sampling. Details
about the calculation are given elsewhere \cite{Paschek:2004:1}.
Figure 3 shows the excess chemical potential for Argon
dissolved in TIP5P-E water for all simulated state points. 
First of all we would like to emphasize that
the solubility increases strongly when penetrating
the deeply supercooled region with $\mu_{ex}$ becoming even negative.
The insert of Figure \ref{fig:03} compares the experimental data
of Kennan and Pollack \cite{Kennan:90} with data from our simulation.
The pressure dependence is expressed by
a positive partial molar volume of $30\,\mbox{cm}^3\,\mbox{mol}^{-1}$, 
close to the experimental value of $25\,\mbox{cm}^3\,\mbox{mol}^{-1}$
\cite{voldetails}. The decreasing solubility upon density increase
basically reflects the loss of available free volume. A larger partial
molar volume in the simulation is hence in line with the increased 
isothermal compressibility observed for the TIP5P-E model, 
indicated by Figure \ref{fig:02}a.
Two regimes of small and strong pressure dependence are denoted, obviously
related to the transformation between the high and low density liquid
forms of water. We would like to point out that we find an increase of
the partial molar volume for Ar around normal pressure for the
$260-255\,\,\mbox{K}$ isotherms.
Kennan and Pollack made a similar observation
for Xenon around $298\,\mbox{K}$. If, and how these two observations are
related has to be further investigated.
Figure 4 shows the temperature dependence of $\mu_{ex}$
for several selected isobars ranging from $-200\,\mbox{MPa}$ to
the supercritical $350\,\mbox{MPa}$. 
The experimental $0.1\,\mbox{MPa}$ isobar \cite{Prini:89} 
is given for comparison. From 
computer simulations of water, S.\ Garde et al. 
have derived an information
theory (IT) model \cite{Garde:96:3,Hummer:2000}, giving simple analytic
expressions for the hydrophobic
hydration as a function of temperature and density. 
The leading term in the IT model strongly suggests a quadratic
relation between the excess chemical potential and the water number density
according to $\mu_{ex}/k\!\approx\!\rho^2 T v^2/2\sigma_n^2$
\cite{Hummer:2000}, where $v$ denotes the volume of 
a hydrophobic hard sphere particle, 
while $\sigma_n^2\!=\!\left< n^2 \right>-\left< n \right>^2$
indicates the variance of the number of water molecules in a
sphere of volume $v$. The lines in Figure 4 indicate a temperature dependence 
as suggested by the IT model, assuming the term 
$v^2/2\sigma_n^2$ to be constant 
and shifting the isobars by a constant
offset to account for attractive interactions.
Well reproduced is the change in slope when passing the
transformation into the low density liquid, qualitatively correct
even for the supercritical isotherm at $350\,\mbox{MPa}$,
suggesting a discontinuity of $\mu_{ex}$. 
Hence an important conclusion is
that the weak monotonous temperature and density dependence of 
$\mu_{ex}$, as proposed by the IT model, and confirmed for
the high temperature regime \cite{Garde:96:3}, is apparently also valid for the 
deeply supercooled region. This seems to be in line with the
continuous structural transformation of water as a function of density 
(Figure \ref{fig:02}c) and
temperature (not shown). The thermodynamics of
hydrophobic hydration of small solutes is 
therefore largely predetermined by the location of
the water isobars in the $\rho,T$-plane. 
As a consequence, also a waterlike liquid showing a continuous  
singularity free transition might exhibit a quite similar 
$\mu_{ex}$-dependence (apart from the
discontinuous behavior of $\mu_{ex}$ above $P_c$). 
As shown
in the insert of Figure 4, the change in slope of $\mu_{ex}$ is related
to extrema in the temperature dependence of the entropy and enthalpy contributions.
The minimum in the enthalpy causes a change in sign of the corresponding 
heat capacity from positive at high temperatures to negative
for the low density liquid. 
Apparently, a prominent signature of the
``hydrophobic hydration'' vanishes while passing the transition at about
$250\,\mbox{K}$. In addition, an increasing slope of $h_{ex}$ upon approaching
$250\,\mbox{K}$, as seen in the insert, and more prominently pronounced with
increasing pressure (not shown), is consistent with the experimentally
observed increase of the solvation 
heat capacity upon cooling \cite{Silverstein:2000}.
Following the interpretation of the hydrophobic hydration
according to Dill et al.\ \cite{Southall:2002}, 
the change in sign  of $C_p$ is explained 
as follows: In the LDL-regime, water's
coordination number approaches four, while the structure corresponds to
a continuous random tetrahedral network. 
Due to the lack of additional possible
hydrogen binding partners, 
the environment of a water molecule in the LDL-bulk and
in the hydrophobic hydration shell has greater similarity
than in the ``high density liquid'' at ambient temperatures. 
This structural convergence in bulk and hydration shell in the LD-liquid 
leads to a smaller entropy difference and hence to an increase in $s_{ex}$.
Moreover, a larger
hydrophobic particle in the LD-liquid might even introduce disorder and
hydrogen bond breaking into the (over) stretched
hydrogen  bond network 
and  thus might even experience a {\em positive}
hydration entropy.
A further investigation on particle 
size dependence, hydrophobic aggregates and hydrophobic 
interaction might reveal a behavior markedly different from that at
ambient conditions. 

This work was supported by the Deutsche Forschungsgemeinschaft
(Forschergruppe 436). I would like to thank Angel
E. Garc\'ia, Ivan Brovchenko and Alfons Geiger for valuable discussions.


\begin{thebibliography}{34}
\bibitem[{VTR()}]{VTREMD}
\bibinfo{note}{State swapping moves between two states $i$ and $j$ are accepted
  with a probability $P_{acc} \mbox{min} \{ 1,\exp[ \beta_i \left(
  U(\vec{s}_i^N;L_i) - U(\vec{s}_j^N;L_i) \right) + \beta_j \left(
  U(\vec{s}_j^N;L_j) - U(\vec{s}_i^N;L_j) \right) ]\}$. Here, $\vec{s}_i^N$
  represents the set of scaled coordinates $\vec{s}_N\!=\!L^{-1}\,\vec{r}_N$ of
  the entire $N$-particle system belonging to state $i$. $U(\vec{s}_i^{N};L_i)$
  denotes the potential energy of configuration $\vec{s}_i^N$ at volume
  $V_i=L_i^3$, whereas $U(\vec{s}_i^{N};L_j)$ represents the configurational
  energy belonging to $\vec{s}_i^N$ at volume $V_j$. Whether a state swapping
  move or an MD move is executed, is chosen at random with a probability of
  $0.2$ for selecting state swapping moves.}
\end{thebibliography}
\end{document}